\documentclass[aps, prd, amsmath, preprintnumbers, floats, floatfix, twocolumn, footinbib, showpacs]{revtex4-1}
\usepackage{graphicx}
\usepackage{color}

\newcommand{\be}{\begin{eqnarray}}
\newcommand{\ee}{\end{eqnarray}}

\begin{document}

\title{Attempt to find a correlation between the spin of stellar-mass black hole candidates and the power of steady jets: relaxing the Kerr black hole hypothesis}

\author{Cosimo Bambi}
\email{Cosimo.Bambi@physik.uni-muenchen.de}
\affiliation{Arnold Sommerfeld Center for Theoretical Physics,\\
Ludwig-Maximilians-Universit\"at M\"unchen, D-80333 Munich, Germany}

\date{\today}

\preprint{LMU-ASC-31-12}

\begin{abstract}
The rotational energy of a black hole can be extracted via the
Blandford-Znajek mechanism and numerical simulations suggest
a strong dependence of the power of the produced jet on the black
hole spin. A recent study has found no evidence for a correlation
between the spin and the power of steady jets. If the measurements
of the spin and of the jet power are correct, it leads one to conclude
that steady jets are not powered by the black hole spin. In this
paper, I explore another possibility: I assume that steady jets are
powered by the spin and I check if current observations can be
explained if astrophysical black hole candidates are not the Kerr black 
hole predicted by General Relativity. It turns out that this scenario might
indeed be possible. While such a possibility is surely quite speculative,
it is definitively intriguing and can be seriously tested when future
more accurate measurements will be available.
\end{abstract}


\maketitle

%%%%%%%%%%%%%%%%%%%%%%%%%%%%%%%

\section{Introduction}

Jets and outflows are a quite common feature of accreting compact objects.
In the case of stellar-mass black hole (BH) candidates in X-ray binary
systems, we observe two kinds of jets~\cite{fbg}. {\it Steady or continuous jets} 
occur in the hard spectral state. {\it Transient or episodic  jets}
appear most significantly when the source switches from the hard to the soft state.
Most efforts so far concentrate on the formation mechanism of steady jets. 
One of the most appealing scenarios to explain the formation of steady jets
is the Blandford-Znajek mechanism~\cite{bz}, in which magnetic fields
threading the BH event horizon are twisted and can extract the rotational
energy of the spinning BH, producing an electromagnetic jet. Numerical
simulations show that this mechanism can be very efficient and depends
strongly on the BH spin~\cite{mck,tnm10,tnm11} (but see also Ref.~\cite{livio}
for different conclusions). At the moment it is not clear if the Blandford-Znajek
mechanism can be responsible for the production of steady jets, and in the
literature there are some controversial results. The spin scenario is surely attractive, 
but still unproved. Recently, there have been some studies investigating if 
there is observational evidence for a correlation between BH spin and jet power 
in current data.

In Ref.~\cite{fgr}, Fender et al. considered (separately) all the spin measurements of 
BH binaries reported in the literature and inferred from the continuum-fitting 
method~\cite{cfm1,cfm2} and the K$\alpha$ iron line analysis~\cite{ref-iron}. 
For steady jets in the hard spectral state, they estimate the
jet power via a normalization $C$, defined by
\be
\log_{10} L_{\rm radio} = C + 0.6 \left( \log_{10} L_{\rm X} - 34 \right) \, ,
\ee
where $L_{\rm radio}$ and $L_{\rm X}$ are, respectively, the radio and X-ray
luminosity of the object. Independently, they also consider an estimate of
the jet power from near-infrared data
\be
\log_{10} L_{\rm NIR} = C + 0.6 \left( \log_{10} L_{\rm X} - 34 \right) \, .
\ee
Their plots clearly show no evidence for a correlation between BH spin and
jet power. They thus conclude that: $i)$ the methods used to estimate the
spin parameter are wrong, and/or $ii)$ the methods used to estimate the
jet power are wrong, and/or $iii)$ there is indeed no relation between BH
spin and jet power.

In Ref.~\cite{nm}, Narayan and McClintock proposed that the Blandford-Znajek 
mechanism is responsible for the formation of transient jets. They 
considered the most recent spin measurements obtained via the continuum-fitting 
method from the Harvard-Smithsonian CfA group, which are supposed
to be more reliable, and used a different proxy for the jet power, the peak radio
flux normalized at 5~GHz, which they claimed to be model independent.
They found a correlation between BH spin $a_*$ and jet power
$P_{\rm jet}$, which is consistent with both the theoretical prediction
\be\label{eq-bz-p}
P_{\rm jet} \propto a_*^2
\ee
obtained in Ref.~\cite{bz} for $a_*^2 \ll 1$ and the more accurate one 
\be
P_{\rm jet} \propto \Omega_{\rm H}^2 \propto a_*^2/r_{\rm H}^2 \, ,
\ee 
where $\Omega_{\rm H}$ and $r_{\rm H}$ are, respectively, the angular frequency and
the radius of the event horizon, found in Ref.~\cite{tnm10} and valid even when 
$a_*$ is quite close to 1. In this case, the measurement of the jet power could be
used to test the geometry of the space-time around stellar-mass
BH candidates, as discussed in Ref.~\cite{b12}. However, numerical studies
of the Blandford-Znajek mechanism show the production of steady jets,
while the origin of transient jets remains unclear and other scenarios may
look more appealing, such as episodic ejection of plasma blobs~\cite{fy}.

In this paper, I explore a more speculative possibility. I assume that
steady jets are powered by the spin of the compact object and that the
method used in Ref.~\cite{fgr} to estimate the jet power is correct. I also 
assume that the continuum-fitting method is a robust technique, but that 
the spin measurements reported in the literature are wrong because the 
BH candidates in X-ray binary systems are not Kerr BHs.

\section{Spin measurements}

A geometrically thin and optically thick accretion disk in a stationary,
axisymmetric, and asymptotically flat space-time can be conveniently
described by the Novikov-Thorne model~\cite{nt}. In a Kerr background,
the thermal spectrum of this disk depends on five parameters: BH mass
$M$, BH spin $a_*$, mass accretion rate $\dot{M}$, viewing angle $i$,
and distance from the observer $d$. If $M$, $i$, and $d$ can be measured
by optical observations, one can fit the thermal spectrum of the disk
to infer $a_*$ and $\dot{M}$. This technique is called the continuum-fitting
method~\cite{cfm1,cfm2} and can be used only for stellar-mass BH
candidates: the disk temperature scales as $M^{-0.25}$, so the peak of
the spectrum is around 1~keV for stellar-mass objects, but falls in the
UV range for the super-massive BH candidates with $M \sim 10^5 -
10^9$~$M_\odot$. In the latter case, the data are not good because  of
dust absorption.

The basic (astrophysical) assumptions of the continuum fitting-method
have been tested and verified by observations and theoretical studies
(see e.g.~\cite{cfm2} and references therein) and the technique seems to
be robust. However, it relies on the fact that BH candidates in X-ray binary
systems are Kerr BHs, which is still to be proved~\cite{review}. If we consider 
an accretion disk around a compact object with mass, spin, and a deformation 
parameter (measuring possible deviations from the Kerr geometry), the
thermal spectrum of the disk depends on six parameters. In this case,
the continuum-fitting method provides an estimate of the mass accretion
rate $\dot{M}$, which is deduced from the intensity of the spectrum in
the low frequency region, where the details of the geometry of the
background are not important, and an estimate of the radiative
efficiency $\eta = 1 - E_{\rm ISCO}$~\cite{bb}, where $E_{\rm ISCO}$
is the specific energy of a particle at the innermost stable circular
orbit (ISCO), which is supposed to be the inner edge of the disk in the
Novikov-Thorne model. It is thus clear that the continuum-fitting
method cannot distinguish a Kerr BH with spin parameter $a_*$ and
radiative efficiency $\eta$ from a non-Kerr object with a different
spin parameter (not necessarily with $|a_*| \le 1$ as for a Kerr BH~\cite{b11b})
but the same radiative efficiency. One the other hand,
a jet powered by the Blandford-Znajek mechanism should still be
correlated with the spin of the compact object. Let us notice that the
existence of the event horizon is not strictly necessary here: for
example, even neutron stars may have spin powered jets -- we just
need magnetic fields anchored on the neutron star. On the basis of 
general arguments, we should expect that the jet power can be 
written in the following form:
\be\label{eq-th-p}
P_{\rm jet} = A_0 + 
\sum_{n = 1}^{+\infty} A_n |a_*|^{2n} \, ,
\ee
where $A_0$ takes into account a possible non-spin contribution ($ A_0 \ge 0$, as 
$P_{\rm jet}$ cannot become negative for $a_* = 0$)
and the other terms are due to (some version of) the Blandford-Znajek mechanism. 
The latter must depend only on even powers of $a_*$, because the direction 
of the spin should not be important (at least at first approximation).

\section{Non-Kerr space-times}

As a non-Kerr background, here I consider the Johannsen-Psaltis (JP)
metric, which was explicitly proposed in Ref.~\cite{jp} to test the
geometry around BH candidates. Such a metric does not satisfy
Einstein's vacuum equations (unlike, for instance, the Manko-Novikov
solution studied in~\cite{bb}), but it can be seen as a simple and useful 
approximation to describe BHs in putative alternative theories of gravity,
whose gravitational force is either stronger or weaker than the one around 
a Kerr BH with the same mass and spin. In Boyer-Lindquist coordinates, the
JP metric is given by the line element
\begin{widetext}
\be\label{eq-jp}
ds^2 &=& - \left(1 - \frac{2 M r}{\Sigma}\right) (1 + h) \, dt^2
+ \frac{\Sigma (1 + h)}{\Delta + a^2 h \sin^2\theta } \, dr^2
+ \Sigma \, d\theta^2 - \frac{4 a M r \sin^2\theta}{\Sigma} (1 + h) \, dt \, d\phi + \nonumber\\
&& + \left[\sin^2\theta \left(r^2 + a^2 + \frac{2 a^2 M r \sin^2\theta}{\Sigma} \right)
+ \frac{a^2 (\Sigma + 2 M r) \sin^4\theta}{\Sigma} h \right] d\phi^2 \, ,
\ee
\end{widetext}
where $a = a_* M $, $\Sigma = r^2 + a^2 \cos^2\theta$,
$\Delta = r^2 - 2 M r + a^2$, and
\be
h = \sum_{k = 0}^{\infty} \left(\epsilon_{2k}
+ \frac{M r}{\Sigma} \epsilon_{2k+1} \right)
\left(\frac{M^2}{\Sigma}\right)^k \, .
\ee
This metric has an infinite number of deformation parameters $\epsilon_i$ and
the Kerr solution is recovered when all the deformation parameters are set to
zero. However, in order to reproduce the correct Newtonian limit, we have to
impose $\epsilon_0 = \epsilon_1 = 0$, while $\epsilon_2$ is strongly
constrained by Solar System experiments~\cite{jp}. Properties and observational
features of the JP space-times have been discussed in Refs.~\cite{b11,jp12,chen}.
In this work, I will only examine the simplest cases where either $\epsilon_3 \neq 0$ 
or $\epsilon_4 \neq 0$, while the rest of the deformation parameters are set
to zero.

\section{Observations}

Let us now consider the objects studied in Ref.~\cite{fgr}, with the most recent
spin measurements obtained from the continuum-fitting method in the case
of a Kerr background. The list of these objects is reported in Tab.~\ref{tab}. The
spin measurements (second column) can be easily translated into radiative
efficiency measurements (third column). One can then determine the spin
parameter of a non-Kerr compact object with a specific deformation parameter
by looking for the system with the same radiative efficiency, as discussed in
Ref.~\cite{b12}. In this work, I do not consider the measurements 
from the K$\alpha$ iron line because this technique has not yet been well 
studied for non-Kerr metrics; in particular, I am not aware of any simple 
rule to translate a spin measurement obtained in the Kerr background to 
an allowed region in the spin parameter-deformation parameter plane.

\begin{figure*}
\begin{center}
\includegraphics[type=pdf,ext=.pdf,read=.pdf,width=8.5cm]{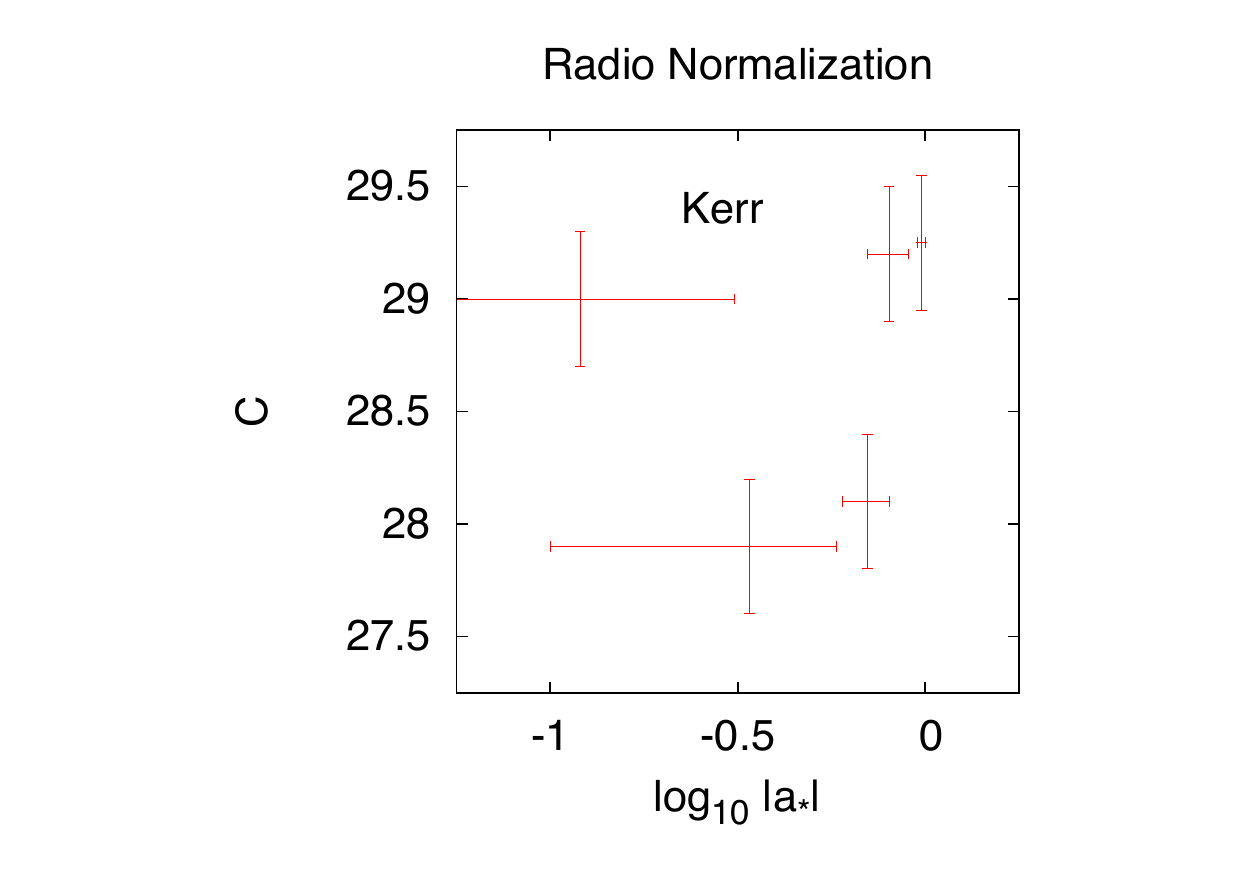}
\includegraphics[type=pdf,ext=.pdf,read=.pdf,width=8.5cm]{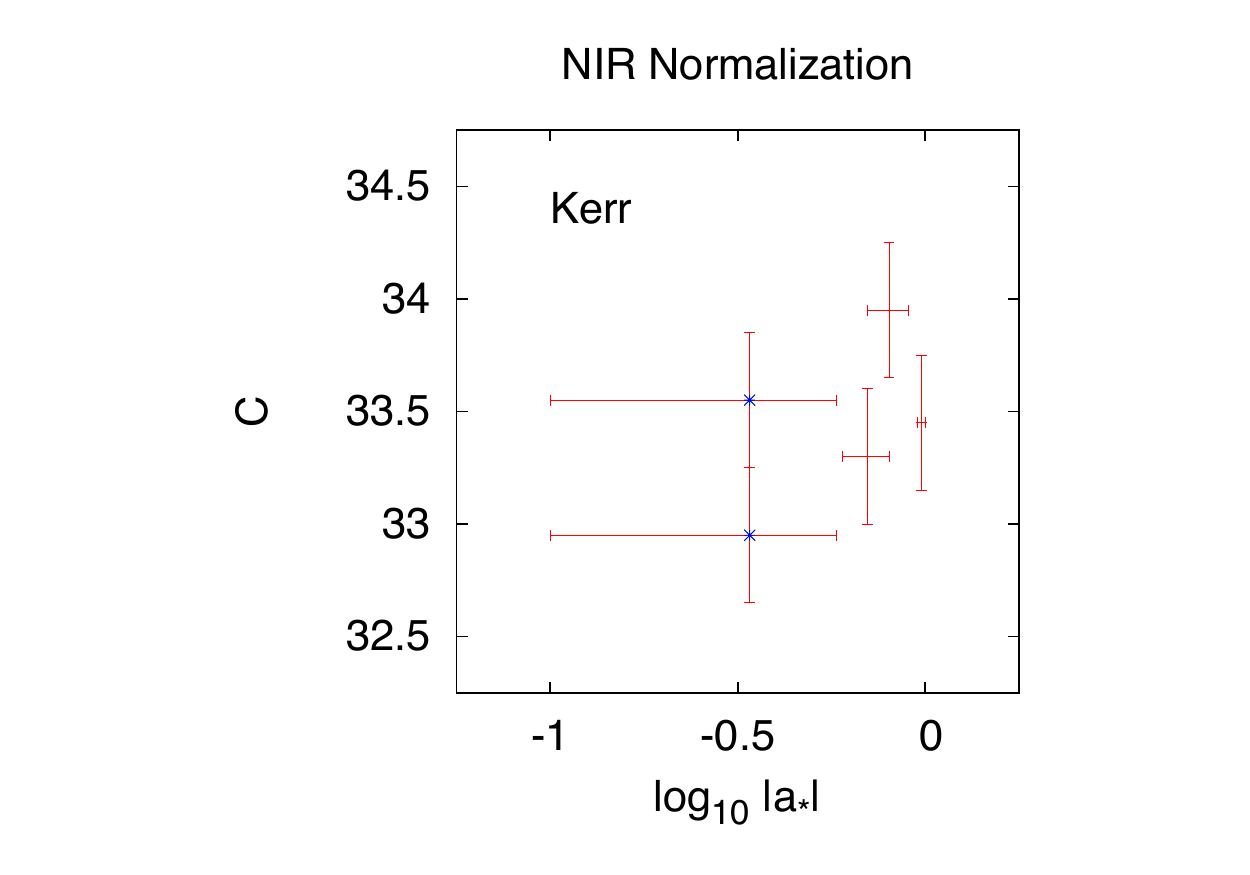}
\vspace{-0.5cm}
\caption{Estimates of the jet power from the radio (left panel) and near-infrared
(right panel) normalizations against the measurements of the BH spin
parameter obtained from the continuum-fitting method and under the
assumption of Kerr background. In the case of the near-infrared normalization,
the object XTE~J1550-564 has two measurements, indicated by the two blue
crosses. The two panels are essentially the right panels of Fig.~4 in Ref.~\cite{fgr}.}
\label{f-kerr}
\end{center}
\end{figure*}

In the case of a Kerr background, the estimates of the jet power via the radio
and near-infrared normalization show no evidence for a correlation with the
spin measurements obtained with the continuum-fitting method, see
Fig.~\ref{f-kerr}. The two panels in Fig.~\ref{f-kerr} are essentially the two
right panels in Fig.~4 of Ref.~\cite{fgr}, with the sole difference being that here I am
using only the most recent measurements of the continuum-fitting method
reported in Refs.~\cite{bh1,bh2,bh3,bh4}. For the objects XTE~J1550-564 and
A0620-00, the spin uncertainty is the one reported in Refs.~\cite{bh3,bh4}.
For GRS~1915+105, 4U~1543-47, and GRO~J1655-40, the spin uncertainty
reported in Refs.~\cite{bh1,bh2} has been doubled, as done, for instance,
in Ref.~\cite{nm}, because the analysis of these objects were performed
a few years ago within a less sophisticated theoretical framework. The
uncertainty in $C$ is (rather arbitrarily) assumed to be 0.3 dex, as in Ref.~\cite{fgr}. 
This normalization as a proxy for the jet power has been criticized as
model-dependent by Narayan and McClintock~\cite{nm}. In the case of
the near-infrared normalization, the object XTE~J1550-564 has two
measurements, indicated by the two blue crosses in the right panel of
Fig.~\ref{f-kerr}, obtained respectively from the rise and from the decay of
an outburst.

Fig.~\ref{f-kerr} clearly shows no evidence for a correlation between BH 
spin and radio/near-infrared normalization. However, in order to be more 
quantitative, especially for the discussion of possible deviations from the 
Kerr geometry, we need to consider a specific theoretical prediction for
the value of $C$ and define a $\chi^2$ which properly takes into account  
the uncertainty in the measurements of the continuum-fitting method
and in the estimate of $C$. 

\vspace{0.5cm}

\begin{table*}
\begin{center}
\begin{tabular}{c c c c c c c c c c c c c}
\hline
\hspace{.5cm} & BH Binary & \hspace{.5cm} & $a_*$ &  \hspace{.5cm} & $\eta^{\rm obs}$ &  \hspace{.5cm} & $C^{\rm obs}$ (Radio) &  \hspace{.5cm} & $C^{\rm obs}$ (NIR) &  \hspace{.5cm} & Reference &  \hspace{.5cm} \\
\hline
& GRS 1915+105 & & $0.975$, $0.95 < a_* < 1$ & & $0.224$, $0.190 < \eta < 0.423$ & & $29.25 \pm 0.3$ & & $33.45 \pm 0.3$ & & \cite{bh1} & \\
& 4U 1543-47 & & $0.8 \pm 0.1$ & & $0.122^{+0.034}_{-0.018}$ & & $29.2 \pm 0.3$ & & $33.95 \pm 0.3$ & & \cite{bh2} & \\
& GRO J1655-40 & & $0.7 \pm 0.1$ & & $0.104^{+0.018}_{-0.013}$ & & $28.1 \pm 0.3$ & & $33.3 \pm 0.3$ & & \cite{bh2} & \\
& XTE J1550-564 & & $0.34 \pm 0.24$ & & $0.072^{+0.017}_{-0.011}$ & & $27.9 \pm 0.3$ & & $32.95 \pm 0.3$ & & \cite{bh3} & \\
& & & & & & & & & $33.55 \pm 0.3$ & & \\
& A0620-00 & & $0.12 \pm 0.19$ & & $0.061^{+0.009}_{-0.007}$ & & $29.0 \pm 0.3$ & & -- & & \cite{bh4} & \\
\hline
\end{tabular}
\end{center}
\vspace{-0.2cm}
\caption{The five stellar-mass BH candidates of which the spin parameter $a_*$ has been
estimated with the continuum-fitting method and for which we can get an estimate of the power
of steady jets. The accretion efficiency in the third column has been deduced from the
corresponding $a_*$ for a Kerr background. The normalizations $C$ in the fourth and
fifth columns have been inferred from Fig.~4 of Ref.~\cite{fgr}.}
\label{tab}
\end{table*}

\begin{figure*}
\begin{center}
\includegraphics[type=pdf,ext=.pdf,read=.pdf,width=8.5cm]{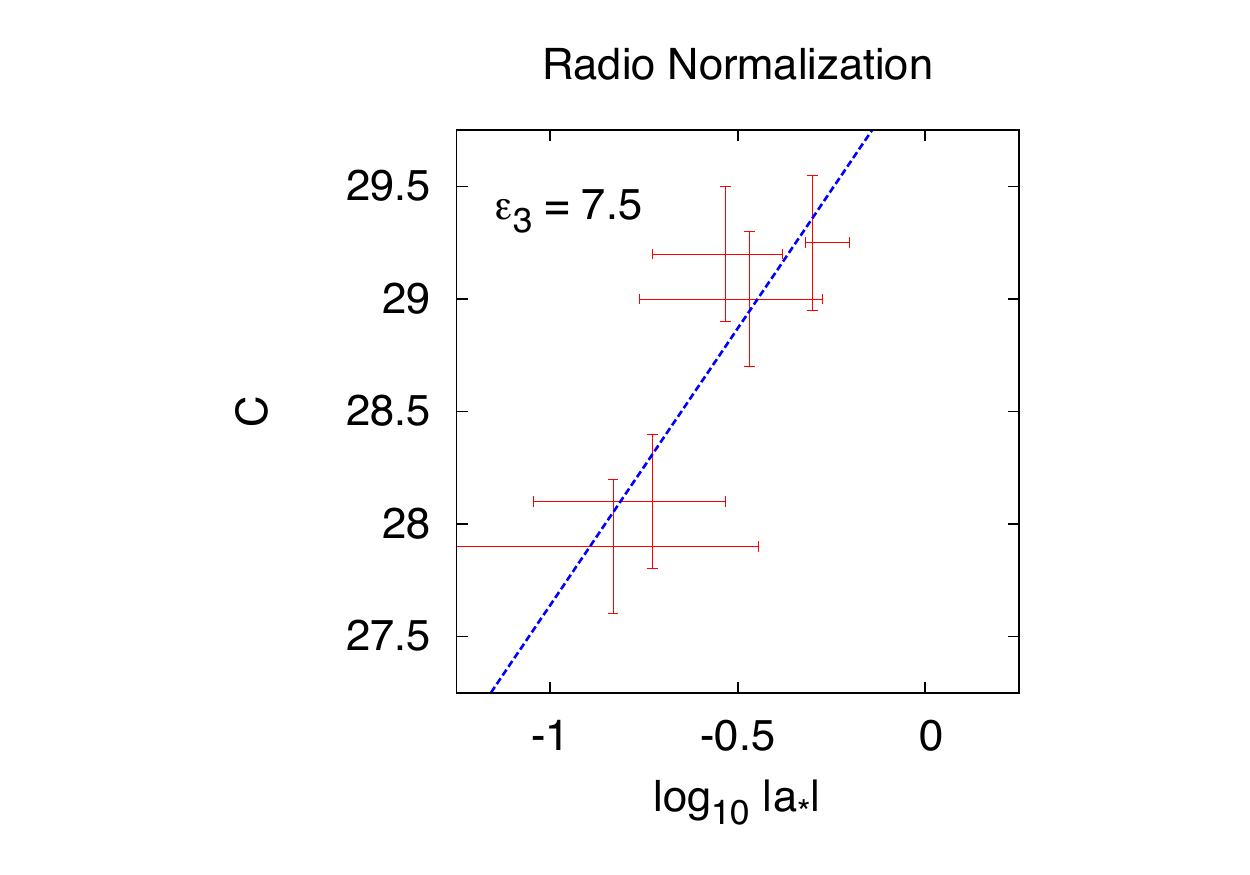}
\includegraphics[type=pdf,ext=.pdf,read=.pdf,width=8.5cm]{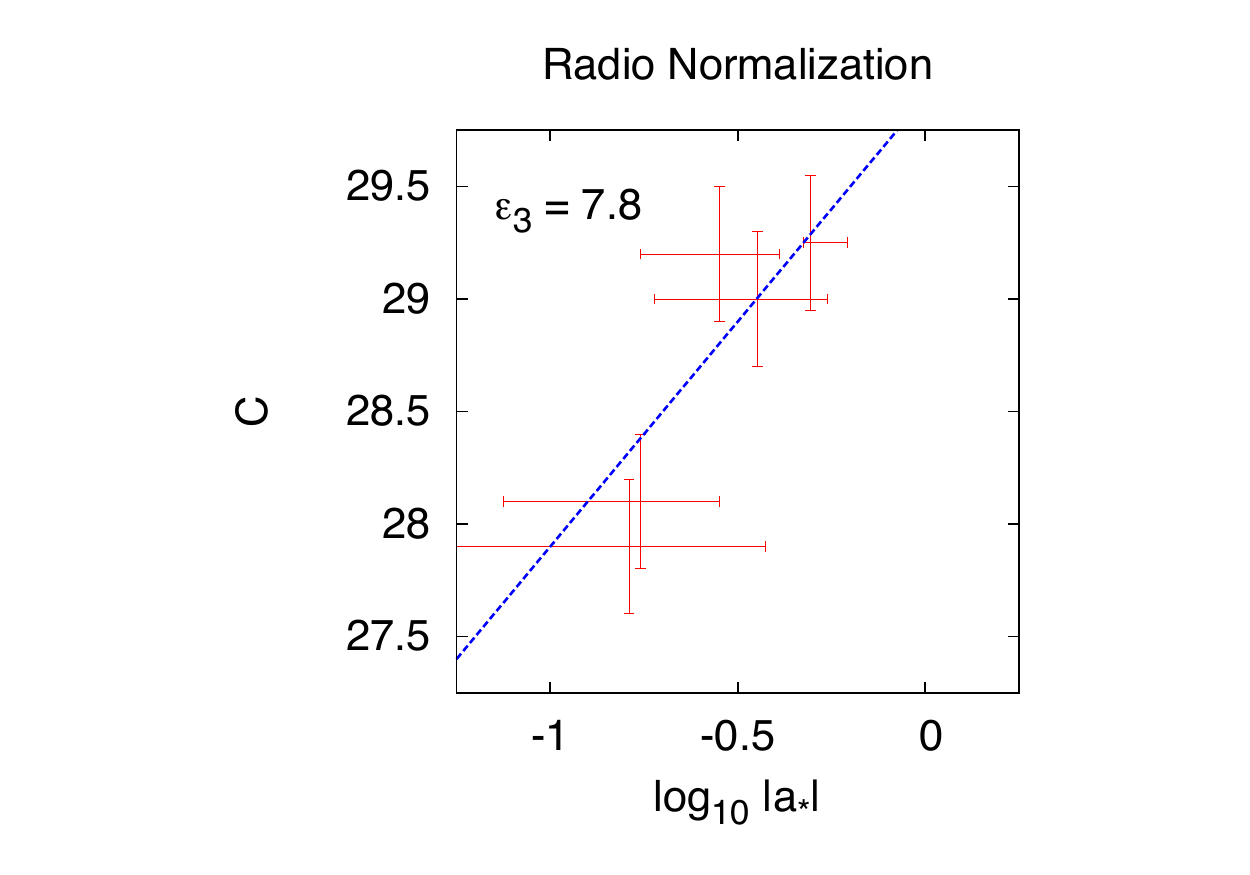}
\vspace{-0.5cm}
\caption{Best fit in the case of the JP background with non-vanishing deformation
parameter $\epsilon_3$, for the jet model 1 (left panel) and the jet model 2 (right 
panel). See text for details.}
\label{f-jp-b}
\end{center}
\end{figure*}

\section{Jet model 1: $P_{\rm jet} = \alpha |a_*|^\beta$}

If we neglect a possible non-spin contribution to the power of the jet,
a simple form for $P_{\rm jet}$ is
\be\label{eq-power}
P_{\rm jet} = \alpha |a_*|^\beta \, ,
\ee
and therefore
\be
C^{\rm th} = \beta \log_{10}|a_*| + \alpha'
\ee
where $\alpha' = \log_{10}\alpha$. $\alpha'$ and $\beta$ are the two 
parameters of the jet model: if we knew all the details of the jet formation, 
they could be theoretically computed, but here they will be 
determined by fitting the data. Let us notice that $\beta$ does not necessarily
need to be an even integer number, because Eq.~(\ref{eq-power})
is a simplified form of Eq.~(\ref{eq-th-p}), in which there are potentially
many terms of the form $(a_*)^{2n}$. $\beta$ should be close to 2 if
the first term is the dominant one, and it should be larger than 2 if
$A_n \neq 0$ for some $n > 1$.

Including the possibility of a (generic) deformation parameter $\epsilon$, the 
$\chi^2$ can be defined as
\begin{widetext}
\be
\chi^2 (\alpha', \beta, \epsilon) = 
\min_{\{ a_{* \, i} \}} \left[ 
\sum_{i = 1}^{N} 
\frac{\left[ C^{\rm th} (\alpha', \beta, a_{* \, i})
- C_i^{\rm obs} \right]^2}{\sigma_C^2} +
\sum_{i = 1}^{N} 
\frac{\left( a_{* \, i} - a_{* \, i}^{\rm obs} \right)^2}{\sigma_i^2} \right] \, ,
\ee
\end{widetext}
where $\sigma_C = 0.3$, $a_{* \, i}^{\rm obs} = a_{* \, i}^{\rm obs} 
(\epsilon, \eta_i^{\rm obs})$, and
\be
\sigma_i =
\begin{cases}
\sigma_i^+ & \text{if } a_{* \, i} > a_{* \, i}^{\rm obs} \, , \\ 
\sigma_i^- & \text{if } a_{* \, i} < a_{* \, i}^{\rm obs} \, 
\end{cases}
\ee
is the uncertainty on $a_{* \, i}^{\rm obs}$ (as $a_{* \, i}^{\rm obs}$
is obtained from $\eta^{\rm obs}_i$, $\sigma_i^+ \neq \sigma_i^-$). 
$\chi^2$ has 3 degrees of freedom, as the spins $\{ a_{* \, i} \}$
are considered as measured quantities. In this and in the next section, 
I will consider only the radio measurements, while the near-infrared 
data will be briefly discussed in Section~\ref{s-dis}. As we have 5 
radio estimates of $C$, $\chi^2_{\rm red} = \chi^2$. If we assume 
the Kerr background, $\chi^2$ has 2 degrees of freedom, and therefore
$\chi^2_{\rm red} = \chi^2/2$.

\subsection{Kerr black holes}

If we assume that the BH candidates are the Kerr BHs predicted by
General Relativity, we have only two fit parameters, $\alpha'$ and 
$\beta$. The best fit is found for
\be
\alpha' &=& 29.3 \, , \nonumber\\
\beta &=& 2.8 \, ,
\ee
with $\min \chi^2_{\rm red} = 6.03$, which clearly confirms there is no 
correlation between jet power and BH spin.

\subsection{JP black holes with $\epsilon_3$ constant \label{sub-s-jp}}

Let us now consider the possibility that the BH candidates in X-ray binary
systems are not necessarily the Kerr BHs predicted by General Relativity,
but that the geometry of the space-time around them can be described by
the JP metric with a deformation parameter. In the case of a background
with arbitrary $\epsilon_3$ and $\epsilon_i = 0$ for $i \neq 3$, one finds
the best fit for
\be
\epsilon_3 &=& 7.5 \, , \nonumber\\
\alpha' &=& 30.1 \, , \nonumber\\
\beta &=& 2.46 \, ,
\ee
with $\min \chi^2_{\rm red} = 0.94$. 
The plot spin vs $C$ for $\epsilon_3 = 7.5$ is shown in the left panel of
Fig.~\ref{f-jp-b}, in which the dashed blue line has slope 2.46. Such a value
for $\beta$ is not far from the theoretical expectation of 2 found in Ref.~\cite{bz}.

Let us notice that the possibility of a non-vanishing deformation parameter
is in contradiction with the finding of Ref.~\cite{b12}, where it was
concluded that $\epsilon_3$ must be small. The assumptions of the two papers 
are indeed different. Here, we assume that {\it steady jets} are powered by the spin
and we try to find the most favorable metric deformation to recover a correlation
between the measured spins and jet powers. At the same time, we do not 
believe that {\it transient jets} originate from the Blandford-Znajek mechanism, 
despite the correlation found by Narayan and McClintock in Ref.~\cite{nm}. 
In Ref.~\cite{b12}, we believed in the correlation found by Narayan and 
McClintock and that it was enough to say that $\epsilon_3$ must be small.
It is important to stress that, even when adopting individual deformation parameters,
it is impossible to reconcile the contradicting claims: the point is that for the
object A0620-00 we have a powerful steady jet and a weak transient jet,
which demand respectively a high and a low value of the spin parameter
if the jet is really powered by the spin.

\subsection{JP black holes with $\epsilon_3 = \gamma a_*^2$ and $\gamma$ constant \label{subs-g}}

The JP space-time is a phenomenological background proposed to describe,
at a first approximation, putative non-Kerr BHs. The deformation parameters are 
used to estimate possible deviations from the Kerr geometry, but their actual 
physical meaning is not clear. In particular, there is no reason to expect that 
the value of the deformation parameters must be the same for all the objects. 
A specific value of $\epsilon_3$ for every BH candidate may sound too arbitrary 
and even not very natural, as it would recall a conserved charge which belonged to
the progenitor star. On the other hand, a deformation parameter depending 
on the spin sounds much more physical. For instance, the lowest order
deviation from the Kerr background of the space-time around a neutron star
is the mass-quadrupole moment, which is thought to be well approximated
by the form~\cite{laa}
\be
Q = - (1 + \xi) a_*^2 M^3 \, , 
\ee
where $\xi$ ($\xi = 0$ for a 
Kerr BH) is a parameter of order 1 which depends on the matter equation of 
state, i.e. on the microphysics. The simplest guess for the form of the
deformation parameter of non-Kerr BHs is
\be\label{eq-gamma}
\epsilon_3 = \gamma a^2_* \, ,
\ee
because the deformation should not depend on the spin orientation and
higher order terms in $a_*$ may be subdominant.

\begin{figure}
\begin{center}
\includegraphics[type=pdf,ext=.pdf,read=.pdf,width=8.5cm]{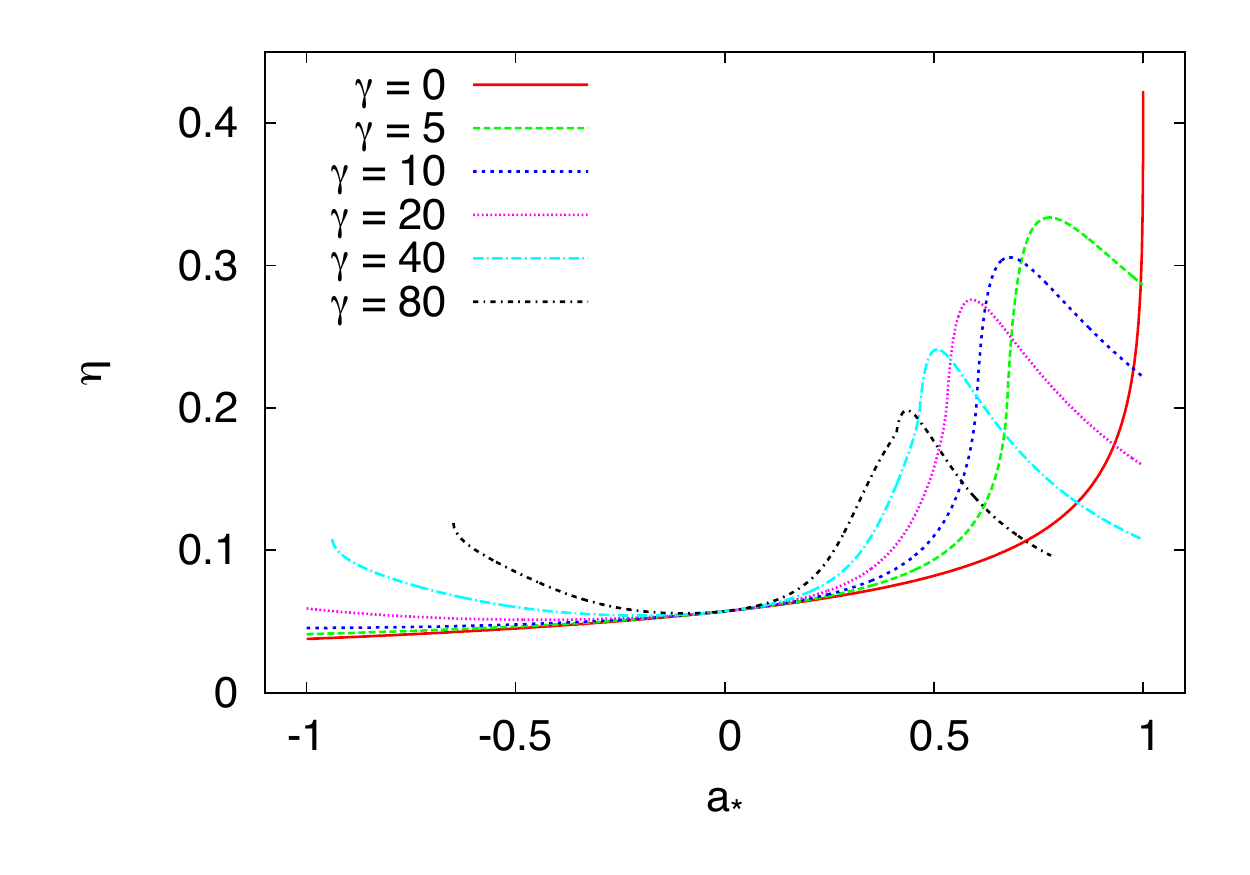}
\vspace{-0.5cm}
\caption{Radiative efficiency $\eta = 1 - E_{\rm ISCO}$ as a function of the 
spin parameter $a_*$ for the JP background with non-vanishing deformation
parameter $\epsilon_3 = \gamma a_*^2$.}
\label{f-eta}
\end{center}
\end{figure}

\begin{figure*}
\begin{center}
\includegraphics[type=pdf,ext=.pdf,read=.pdf,width=8.5cm]{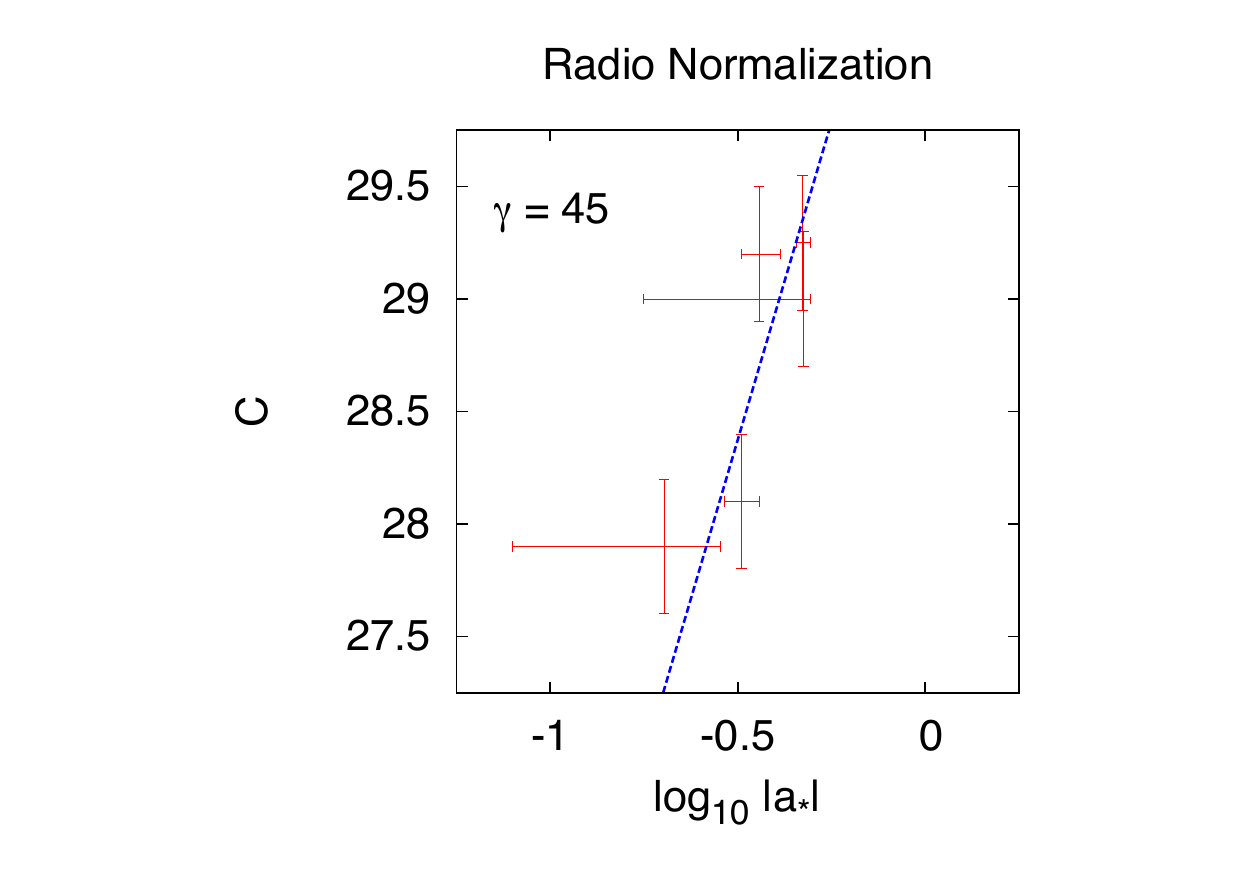}
\includegraphics[type=pdf,ext=.pdf,read=.pdf,width=8.5cm]{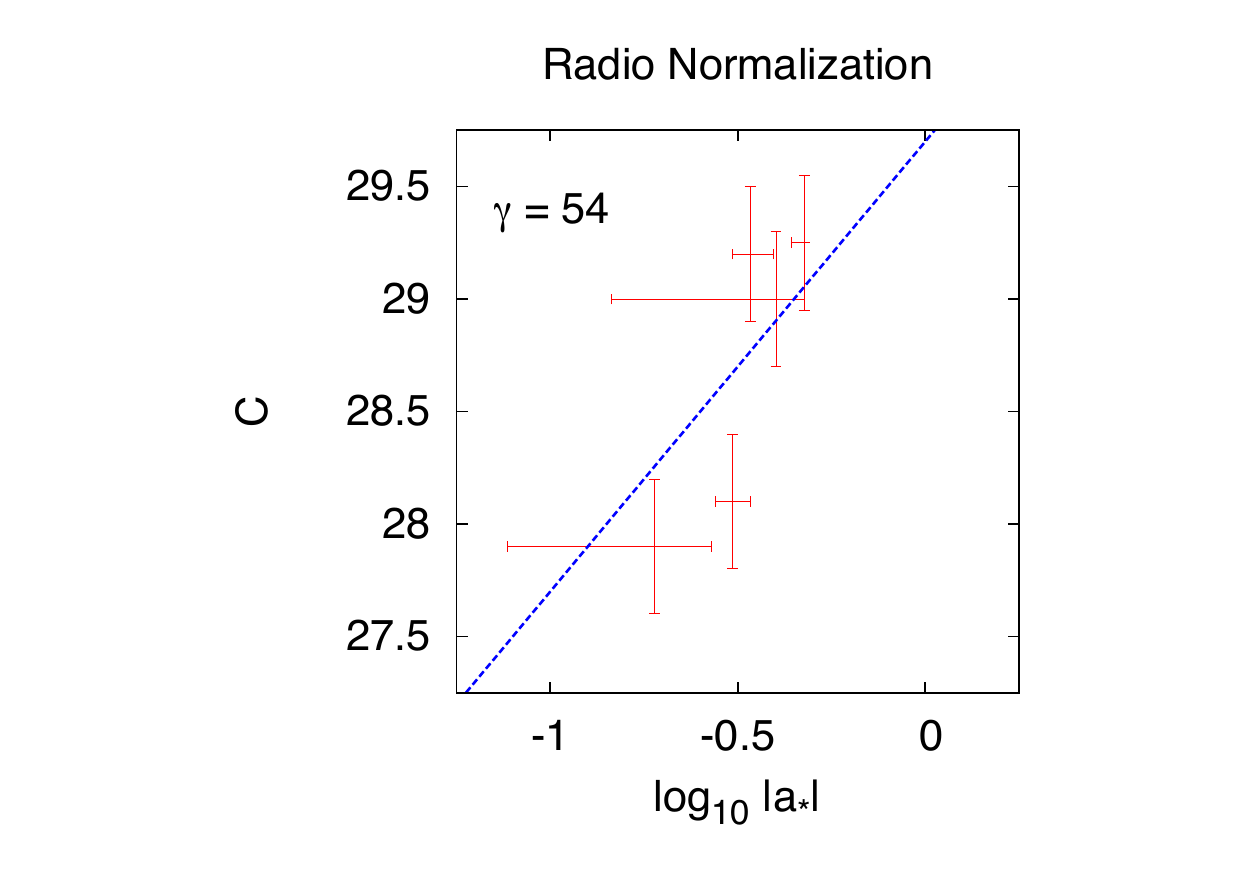}
\vspace{-0.5cm}
\caption{Best fit in the case of the JP background with non-vanishing deformation
parameter $\epsilon_3 = \gamma a_*^2$, for the jet model 1 (left panel) and the 
jet model 2 (right panel). See text for details.}
\label{f-jp-c}
\end{center}
\end{figure*}

One can then repeat the same procedure with the fit parameters $\alpha'$,
$\beta$, and $\gamma$. However, some caution is necessary here. In the
case of Kerr BHs, the continuum-fitting method provides a unique estimate
of the spin parameter of the compact object because there is a one-to-one
correspondence between the radiative efficiency $\eta$ and $a_*$. This
is not longer true in a Kerr background with an arbitrary value of $a_*$;
for instance, for every Kerr BH there is a Kerr naked-singularity (i.e. with 
$|a_*| > 1$) with the same value of $\eta$. However, Kerr-naked singularities
may be excluded for theoretical reasons, as they are apparently impossible 
to produce and, even if created, they would be very unstable~\cite{bara}.
As discussed in Ref.~\cite{b12}, the same conclusions may be true for
JP BHs with a constant $\epsilon_3$. However, this does not seem to be 
true if we assume a constant $\gamma$ and $\epsilon_3$ given by Eq.~(\ref{eq-gamma}). 
The radiative efficiency $\eta = 1 - E_{\rm ISCO}$ as a function of the spin
parameter $a_*$ is shown in Fig.~\ref{f-eta} for some values of $\gamma$.
These BHs can potentially be spun up by the accreting material (for instance, the
envelope of the progenitor star), and then have a counterrotating disk
(formed by the gas coming from the stellar companion). It turns out that 
some values of $\eta$ are common to two configurations with different
spin. In the computation of $\chi^2$, I thus include both the possibilities.
Actually, this problem exists only for A0620-00 when $\gamma > 42$.
The minimization of $\chi^2$ requires
\be
\gamma &=& 45 \, , \nonumber\\
\alpha' &=& 31.2 \, , \nonumber\\
\beta &=& 5.65 \, ,
\ee
and $\min \chi^2_{\rm red} = 2.60$. In this case, $\beta$ is significantly
larger than 2. The best fit is shown in the left panel of Fig.~\ref{f-jp-c}.

\subsection{JP black holes with $\epsilon_4$ constant}

It may be instructive to see what happens if we consider a different deformation
from the Kerr background. If we take the deformation parameter $\epsilon_4$
to be variable and we assume that all the others vanish, we get
\be
\epsilon_4 &=& 18.6 \, , \nonumber\\
\alpha' &=& 30.3 \, , \nonumber\\
\beta &=& 2.16 \, ,
\ee
with $\min \chi^2_{\rm red} = 1.36$. Even in this case, observations would
require a compact object more prolate than a Kerr BH with the same mass
and spin (as $\epsilon_4 > 0$) in order to be consistent 
with the Blandford-Znajek scenario. The best fit is shown in the left panel 
of Fig.~\ref{f-jp-d}.

As long as we consider a single deformation parameter, the effects 
produced by $\epsilon_3$ or by any $\epsilon_i$ with $i > 3$ is very similar. 
This point can be quickly understood by having a look at Figs.~2 and 4 in
Ref.~\cite{b111}, which show the behavior of the radiative efficiency $\eta$ as 
a function of the spin for $\epsilon_3$, $\epsilon_4$ and $\epsilon_5$. 
Actually, the situation is even more general, and the same qualitative 
features can be found by plotting the same figure for the Manko-Novikov 
space-time with a single deformation parameter. In the case of two or more 
non-vanishing deformation parameters, the picture is more complicated. 
If these parameters produce similar deformations (e.g. JP parameters that 
are all positive or all negative), we should still expect the same effects. 
Otherwise, different parameters may produce opposite deformations that 
compensate one another, and it is not easy to predict what may happen. 
In Ref.~\cite{b12}, the case of $\epsilon_4 \neq 0$ was not considered. 
However, on the basis of the above considerations (see also the discussion
in Subsection~\ref{sub-s-jp}), the claim of the present paper and the one 
we can obtain from the transient jets discussed by Narayan and McClintock
are necessarily in contradiction, and the choice of a different $\epsilon_i$
cannot solve the incompatibility.

\section{Jet model 2: $P_{\rm jet} = \alpha |a_*|^2 + \beta$}

In this section, we explore the possibility that the jet power also receives
a contribution from another source of energy and therefore can
be written as
\be
P_{\rm jet} = \alpha |a_*|^2 + \beta \, ,
\ee 
where the contribution from the spin of the compact object is assumed
to be proportional to $|a_*|^2$ because it is the one expected in the
Blandford-Znajek scenario and, even if originally obtained in the limit
$a_*^2 \ll 1$, it is thought to be a good approximation even when
$a_*$ is not very close to 1. The theoretical proxy $C$ is
\be
C^{\rm th} = 2 \log_{10}\left( |a_*| + \beta' \right) + \alpha'
\ee
where $\alpha' = \log_{10}\alpha$ and $\beta' = \sqrt{\beta/\alpha}$.

\subsection{Kerr black holes}

In the case of the Kerr background, we have to fit only the two
parameters of the jet model. The best fit has
\be
\alpha' &=& 29.2 \, , \nonumber\\
\beta' &=& 0.022 \, ,
\ee
with $\min \chi^2_{\rm red} = 6.06$. As was already made clear in
Fig.~\ref{f-kerr}, there is no correlation between measured spins and
jet powers and the model with a non-spin contribution cannot fix
the absence of a correlation.

\subsection{JP black holes with $\epsilon_3$ constant}

The minimization of $\chi^2$ with the jet model 2 in the JP
space-time with constant $\epsilon_3$ suggests the following values
for the fit parameters 
\be
\epsilon_3 &=& 7.8 \, , \nonumber\\
\alpha' &=& 29.9 \, , \nonumber\\
\beta' &=& 0.000 \, ,
\ee
and $\min \chi^2_{\rm red} = 1.05$. It seems like a possible non-spin
contribution to the jet power is not necessary. 
The best fit is shown in the right panel of Fig.~\ref{f-jp-b}.
Let us notice that here, as well as for the other non-Kerr cases discussed
in the next subsections, a negative $\beta'$ would provide a better fit:
for instance, in the present case one finds $\min \chi^2_{\rm red} = 0.82$
for $\epsilon_3 = 7.4$, $\alpha' = 30.1$, and $\beta' = -0.06$.
$\beta' < 0$ could be possible in the presence of a mechanism suppressing
the formation of the jet powered by the spin. However, it cannot
really be a constant independent of the spin, as otherwise we would
obtain $P_{\rm jet} < 0$ for a non-rotating object.

\subsection{JP black holes with $\epsilon_3 = \gamma a_*^2$ and $\gamma$ constant}

For a deformation parameter $\epsilon_3 = \gamma a_*^2$, the new jet
model requires
\be
\gamma &=& 54 \, , \nonumber\\
\alpha' &=& 29.7 \, , \nonumber\\
\beta' &=& 0.000 \, ,
\ee
and $\min \chi^2_{\rm red} = 5.9$. The fit is significantly worse than 
the others, because a correlation is possible with a very strong dependence
of $P_{\rm jet}$ on the spin (as found in Subsection~\ref{subs-g}) and
the possibility of a non-spin contribution is not helpful.
The best fit is shown in the right panel of Fig.~\ref{f-jp-c}.

\subsection{JP black holes with $\epsilon_4$ constant}

\begin{figure*}
\begin{center}
\includegraphics[type=pdf,ext=.pdf,read=.pdf,width=8.5cm]{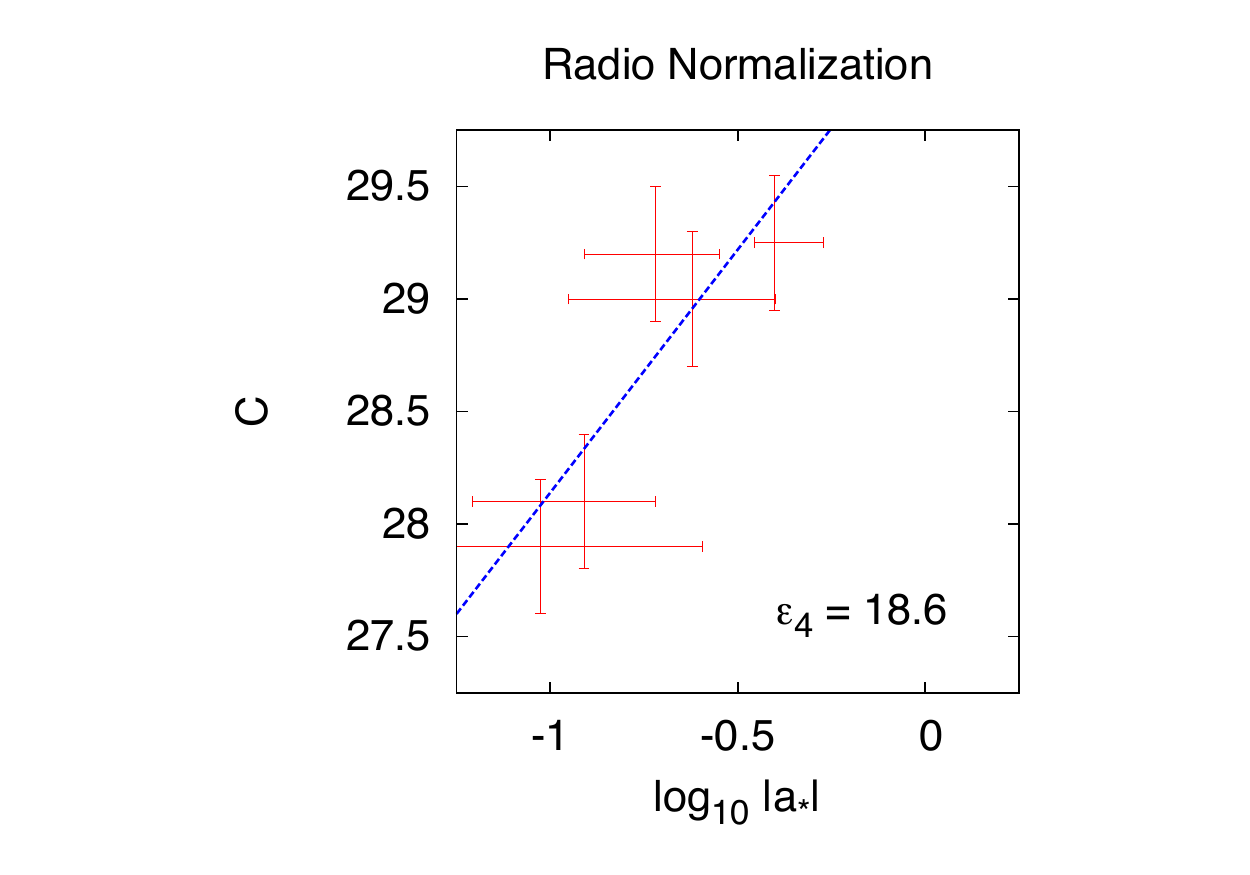}
\includegraphics[type=pdf,ext=.pdf,read=.pdf,width=8.5cm]{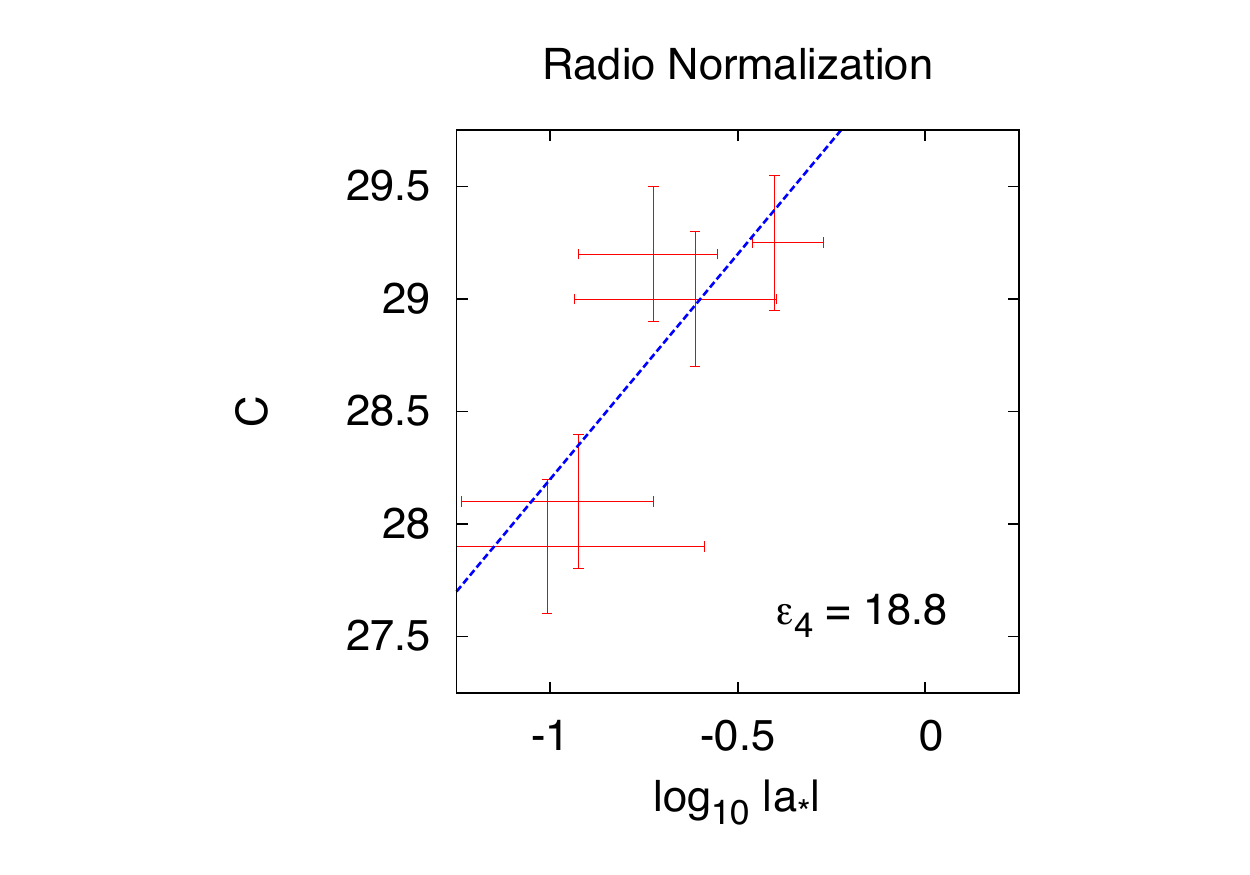}
\vspace{-0.5cm}
\caption{Best fit in the case of the JP background with non-vanishing deformation
parameter $\epsilon_4$, for the jet model 1 (left panel) and the jet model 2 (right 
panel). See text for details.}
\label{f-jp-d}
\end{center}
\end{figure*}

Lastly, we examine the jet model 2 in the JP background with
constant $\epsilon_4$. In this case, the result is
\be
\epsilon_4 &=& 18.8 \, , \nonumber\\
\alpha' &=& 30.2 \, , \nonumber\\
\beta' &=& 0.000 \, ,
\ee
with $\min \chi^2_{\rm red} = 1.37$. As we found for the JP
space-time with constant $\epsilon_3$, the data do not require
any non-spin contribution to the power of the jets. The best fit
is shown in the right panel of Fig.~\ref{f-jp-d}.

\section{Discussion \label{s-dis}}

The correlation between spin measurements and power estimates of steady
jets found for a non-vanishing deformation parameter, while absent in the
Kerr background, can be easily understood as follows. The continuum-fitting
method provides an estimate of the radiative efficiency $\eta$ and, for a
given deformation parameter, there is a one-to-one correspondence between
$\eta$ and $a_*$: $\eta$ is low for a rapidly-rotating object and a counterrotating
disk ($a_*$ negative) and increases as the spin parameter $a_*$ increases.
In the Kerr background, all the measurements are consistent with a corotating
disk, i.e. $a_* > 0$. In a background with a weaker gravitational force (in the JP
metric, $\epsilon_3$ and $\epsilon_4 > 0$), we find the same radiative efficiency
for objects with $a_*$ lower than that of a Kerr BH. In this case, the BH
candidate A0620-00 is interpreted as a fast-rotating object with a counterrotating
disk. On the contrary, the jet power should be independent, at least at first
approximation, of the spin orientation. This difference is enough to find a
correlation between spin and jet power in current data. This is the only possibility
to have a correlation between spins and jet powers, as A0620-00 has a
low radiative efficiency and a powerful steady jet.

From an astrophysical point of view, the possibility of the existence of a 
fast-rotating object with retrograde spin may be challenging. The continuum-fitting
method requires that the disk is perpendicular to the BH spin; if this assumption
is not fulfilled, the technique is essentially unusable (see e.g. Ref.~\cite{fragile}).
If the binary system formed from the collapse of the same cloud (rather then
after the capture of one of the objects by the other), the disk is indeed expected
on the equatorial plane of the BH, but the spin would more likely be parallel,
not antiparallel, to the angular momentum of the disk. If for some reason the 
disk is not initially on the equatorial plane of the system, the action of the 
Bardeen-Petterson effect can force the inner part of the disk to align on the 
BH spin. However, the possibility of retrograde disks cannot be
excluded a priori~\cite{cfm2}. Retrograde disks have been proposed, for 
instance, to explain the radio-loud/radio-quiet dichotomy of AGNs~\cite{garofalo}.

The discussion of the possibility of a correlation between jet power and spin
in the presence of a suitable deviation from the Kerr geometry has been done
considering only the radio data. Is it possible to find a correlation between
jet power and spin in the case of the near-infrared data? The answer is no,
at least in the case of simple deformations in which there is a one-to-one
relation between $\eta$ and $a_*$. In the case of the radio normalization,
the main problem for a correlation between jet power and spin is the powerful
jet of A0620-00, whose spin would be near 0 in the Kerr space-time. If we
consider only the other 4 measurements, the situation is not so bad. 
In the case of a non-Kerr background, one can explain A0620-00 as a fast
rotating object with a retrograde disk and better adjust the other 4 objects.
Eventually, a correlation is possible. In the case of the near-infrared data,
4U~1543-47 shows a powerful jet, while the ones of objects with higher and
lower radiative efficiency seem to be weaker. Since we have only 4 
measurements (or 3, as the estimate of $C$ for XTE~J1550-564 is ambiguous),
every point is very important and it is not possible to find a correlation
between jet power and spin with the same trick used for the radio data.

Lastly, as already stressed at the end of Subsection~\ref{sub-s-jp}, 
the possibility of a correlation between spin and the power of steady jets in
a non-Kerr background cannot be compatible with the one between spin and 
the power of transient jets found by Narayan and McClintock in Ref.~\cite{nm},
as the latter exists only in Kerr space-times and disappears as the deformation 
parameter $\epsilon_3$ increases~\cite{b12}.

\section{Conclusions}

While there are indications suggesting that steady jets in the hard
spectral state may be powered by the BH spin, the study reported
in Ref.~\cite{fgr} shows that there is no evidence for a correlation between
spin measurements and the power of steady jets in BH X-ray binaries. This leads one to
conclude that: $i)$ the spin measurements are wrong, and/or $ii)$ the estimates
of the jet power are wrong, and/or $iii)$ there is indeed no strong relation
between BH spin and jet power. In this paper, I explored the first
possibility, focusing only on the most recent
measurements obtained from the continuum-fitting
method~\cite{bh1,bh2,bh3,bh4}. A key-ingredient of the standard
approach is the assumption of the Kerr BH hypothesis; that is, the
stellar-mass BH candidates in X-ray binaries must be the Kerr BH predicted
by General Relativity. If the BH candidates in X-ray binaries are not
Kerr BHs, the continuum-fitting method provides a wrong estimate
of the spin parameter. I thus investigated if one can find a correlation between
spin measurements and estimates of the jet power in the case that the space-time
around these objects deviates from the Kerr geometry. It turns out that such
a speculative idea might indeed be possible, as shown in Figs.~\ref{f-jp-b},
\ref{f-jp-c}, and \ref{f-jp-d}.
While the current sample of data consists of a small number of objects
with too large uncertainty in the spin measurements and jet power estimates,
the scenario is definitively intriguing and it can be more seriously tested
when future more accurate data will be available.

%%%%%%%%%%%%%%%%%%%%%%%%%%%%%%%

\begin{acknowledgments}
I wish to thank Shantanu Desai and Luca Maccione for useful comments 
and suggestions. This work was supported by the Humboldt Foundation.
\end{acknowledgments}

%%%%%%%%%%%%%%%%%%%%%%%%%%%%%%

\end{document}